\documentclass[superscriptaddress,reprint, nofootinbib]{revtex4-1}  %two column
\RequirePackage{lineno}

\usepackage{amsmath}    % Fancy math fonts
\usepackage{graphicx}   % Include figure filesxr
\usepackage{dcolumn}    % Align table columns on decimal point
\usepackage{bm}         % Bold math
\usepackage{lineno}     % Line numbering package
\usepackage{hyperref}   % Hyperlinks in references
\usepackage{xspace}     % Properly inserts spaces for ath-mode definitions
\usepackage{pdfpages}
\usepackage{subfig}
\begin{document}

\title{$<X_{max}>$ Uncertainty from Extrapolation of Cosmic Ray Air Shower Parameters.}

\author{R.U.~Abbasi}
\affiliation{High Energy Astrophysics Institute and Department of
Physics and Astronomy, University of Utah, Salt Lake City, Utah, USA}

\author{G.B.~Thomson} 
\affiliation{High Energy Astrophysics Institute and Department of
Physics and Astronomy, University of Utah, Salt Lake City, Utah, USA}


\begin{abstract}
%
%\section{Abstract}

Recent measurements at the LHC of the p-p total cross section have reduced the uncertainty in simulations of cosmic ray air showers. In particular of the depth of shower maximum, called $X_{max}$.  However, uncertainties of other important parameters, in particular the multiplicity and elasticity of high energy interactions, have not improved, and there is a remaining uncertainty due to the total cross section.  Uncertainties due to extrapolations from accelerator data, at a maximum energy of $\sim$ one TeV in the p-p center of mass, to 250 TeV ($3\times10^{19}$ eV in a cosmic ray proton's lab frame) introduce significant  uncertainties in predictions of  $<X_{max}>$.  In this paper we estimate a lower limit on these uncertainties.  The result is that the uncertainty in $<X_{max}>$ is larger than the difference among the modern models being used in the field.  At the full energy of the LHC, which is equivalent to $\sim 1\times10^{17}$ eV in the cosmic ray lab frame, the extrapolation is not as extreme, and the uncertainty is approximately equal to the difference among the models.

\end{abstract}
\maketitle

\section{Introduction}

In the study of ultrahigh energy cosmic rays one area of great interest is measurements of cosmic ray composition.  Since the flux is too low for balloon-borne or satellite experiments to detect primary cosmic rays, these events are studied by observing the extensive air showers the primaries initiate.  Perhaps the best method of determining composition is measuring the depth of shower maximum.  Cosmic ray protons differ from iron in the mean depth of shower maximum, $<X_{max}>$, by about 80 g/cm$^{2}$ (protons have shower maxima deeper in the atmosphere).  Intermediate weight nuclei are distributed between the proton and iron depths in a logarithmic manner.  The three experiments with the best data on $<X_{max}>$, the High Resolution Fly’s Eye (HiRes)~\cite{Abbasi:2009nf}, the Telescope Array (TA)~\cite{Abbasi:2015bha}, and the Pierre Auger Observatory (PAO)~\cite{Abraham:2010yv,Aab:2014aea}, all agree within their systematic uncertainties in the actual measurement of $<X_{max}>$ as a function of energy~\cite{Abbasi:2015xga} (although there may be differences in their measurements of the RMS of the Xmax distribution ).  The question now becomes how to interpret these measurements in terms of the identity of the primary cosmic rays.  Are they protons or nuclei?

\begin{figure}[!hp]
 \centering
 \includegraphics[width=0.45\textwidth]{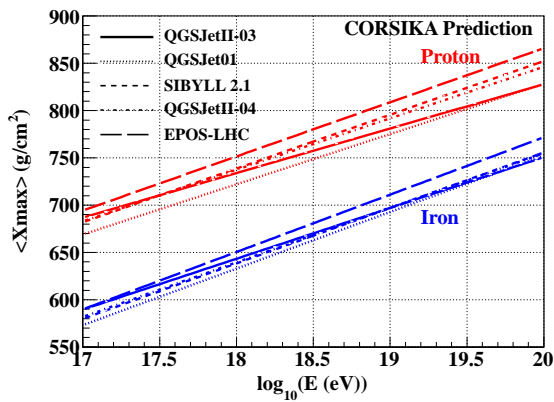}
   \caption{ $<X_{max}>$ predictions from five hadronic models as a function of cosmic ray energy. Red lines are protons while blue lines are iron.}
   \label{fig:elong}
\end{figure}

Up to now the interpretation has been done by comparing the measured $<X_{max}>$ and RMS(Xmax) values to models of cosmic ray air showers.  In these models routines called hadronic generators feed the information about particle interactions to a parent program that keeps track of particles produced in the shower.  The typical parent programs in use are called CORSIKA~\cite{Heck:1998vt} or CONEX~\cite{Bergmann:2006yz,Pierog:2004re,Bossard:2000jh}, and the hadronic generators in use are three version of QGSJet (QGSJet01c~\cite{Kalmykov:1997te}, QGSJetII-3~\cite{Ostapchenko:2005nj}, and QGSJetII-4~\cite{Ostapchenko:2010vb}), Sibyll2.1~(\cite{Fletcher:1994bd},\cite{Engel:1999db}), and EPOS-LHC~\cite{Pierog:2013ria}.  Figure~\ref{fig:elong} shows the predictions of CORSIKA and these five hadronic generators for $<X_{max}>$ of protons and iron as a function of energy.  At $10^{19.5}$ eV the models’ predictions vary by about 35 g/cm$^{2}$, and at $10^{17}$ eV the variation is about 25 g/cm$^{2}$.

The information about particle interactions used in the simulations is based on accelerator measurements where possible, but for cosmic rays at particle energies above those of man-made accelerators extrapolations of accelerator measurements must be made.  For the prediction of $<X_{max}>$ there are  parameters of particle interactions to which the models are particularly sensitive:  the total cross section, the multiplicity of produced particles, the fraction of energy carried away by the leading particles, which is called the elasticity, and the ratio of neutral to charged pions created in the interaction.  Ulrich and collaborators~\cite{Ulrich:2010rg} have tabulated the sensitivity of $<X_{max}>$ predictions for the proton showers at $10^{19.5}$eV by the Sibyll model to these parameters.  The most sensitive is the cross section with a sensitivity of $\triangle X_{max}$/$\triangle$f(E) $\approx$ -100 g/cm$^{2}$ (the minus sign indicates that when the cross section increases $<X_{max}>$ decreases), and the multiplicity, elasticity, and the neutral pion ratio are all about the same, at a value of $\mid \triangle X_{max}$/$\triangle$ f(E) $\mid \approx$  30-40 g/cm$^{2}$. Here f(E) is the fractional change in the parameter at a primary energy of E(eV). f(E) is defined as follows:
\begin{equation}
f(E) = 1 + (f_{19} - 1) \frac{\log(E/10^{15}eV)}{\log(10^{19}eV/10^{15}eV)}
\label{eqn:fe}
\end{equation}

Where E is the shower energy and $f_{19}$ is the scaling factor of the cosmic ray showers at  $10^{19}$ eV. Note that, at $10^{15}$ eV, f(E) $=$ 1. This reflects the fact that the high energy cosmic ray shower simulation models are tuned to the TeVatron. 

\begin{figure}[]
  \centering
  \subfloat{\includegraphics[width=0.45\textwidth]{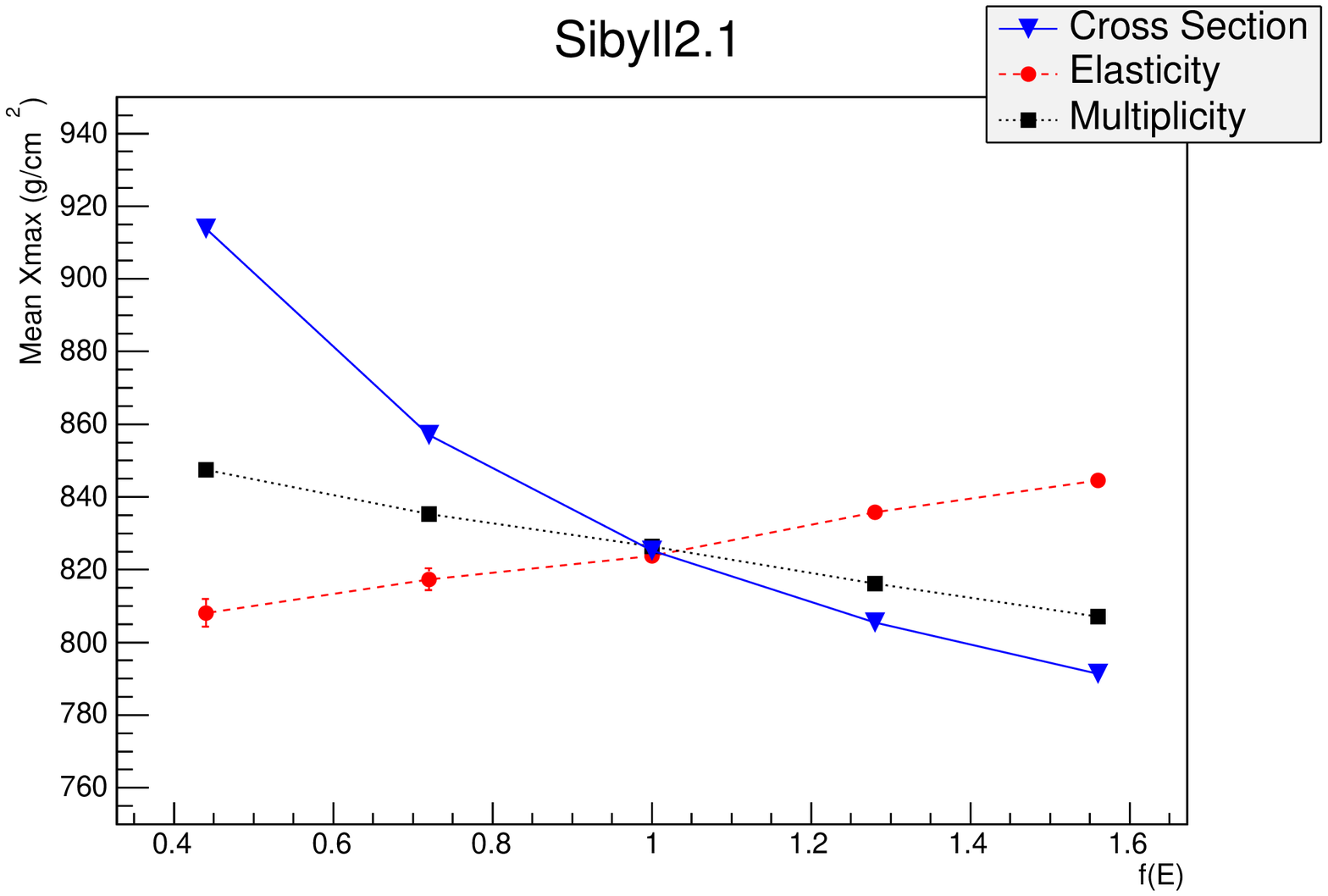}}
  \hfill
  \subfloat{\includegraphics[width=0.45\textwidth]{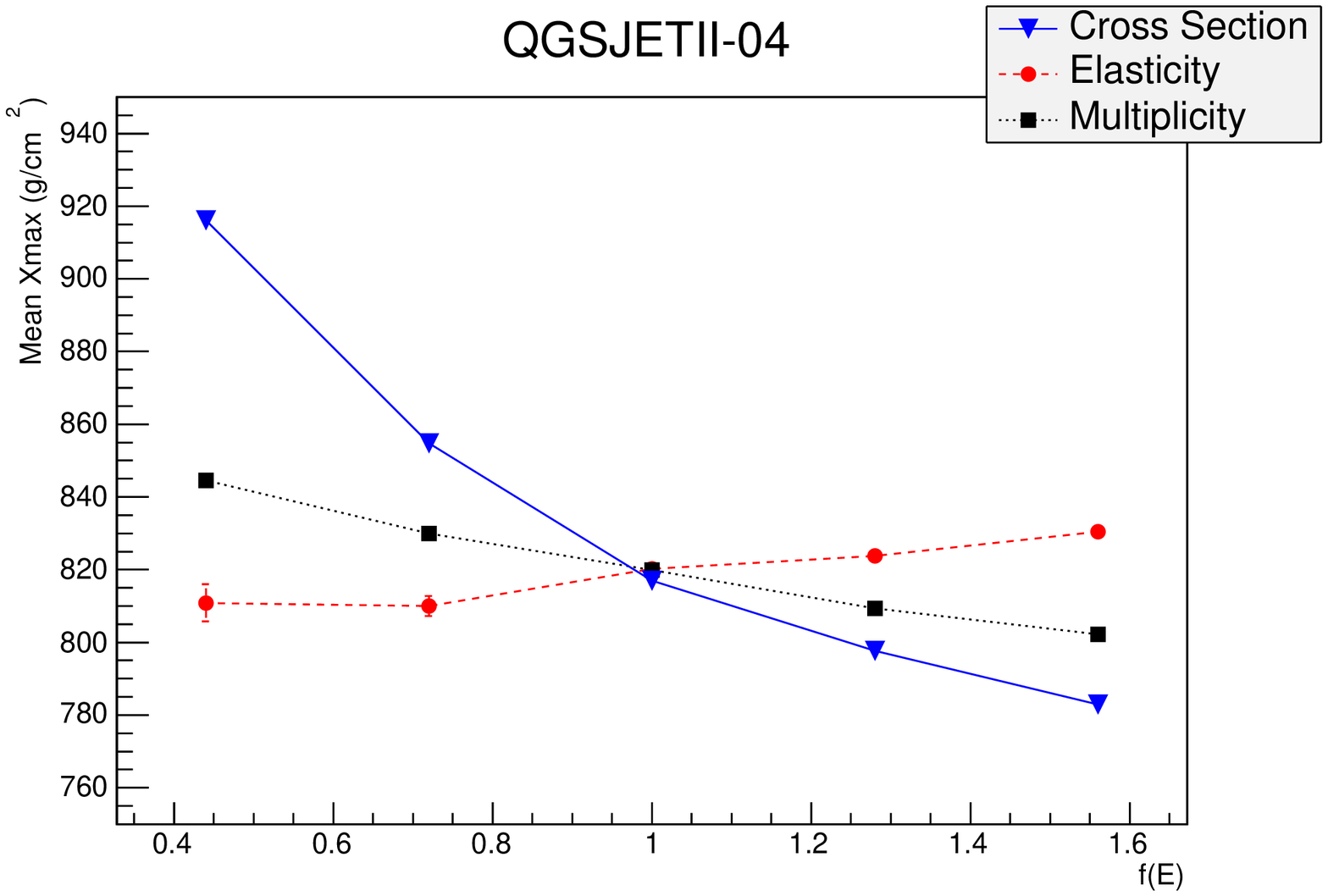}}
  \caption{Dependence of $<X_{max}>$ on cross section, elasticity, and multiplicity at an energy of $10^{19.5}$ eV.  The left panel is the result for Sibyll 2.1, and the right panel is for QGSJetII-4. The lines are to guide the eye. }
  \label{fig:fE}
\end{figure}

We have repeated the Ulrich {\it et al.} estimate, using CONEX~\cite{Bergmann:2006yz,Pierog:2004re,Bossard:2000jh} \footnote[1]{ Revision conex4.44}, for the proton showers for four models and two energies:  the models we used are Sibyll 2.1, QGSJet01c, QGSJetII-4, and EPOS-LHC; evaluated at the energies $10^{17}$ and $10^{19.5}$ eV.  The Sibyll and QGSJetII-4 results at $10^{19.5}$ eV are shown in Figure~\ref{fig:fE}.  Our Sibyll 2.1 calculations reproduce those of Ulrich  {\it et al.} .

In examining the extrapolation of accelerator measurements, the widest energy ranges are available for p-p interactions, but the relevant quantities are needed for p-air interactions.  Taking the multiplicity as an example, a nucleus amplifies differences in multiplicity so extrapolating p-p measurements is a conservative method for learning about uncertainties in the depth of shower maximum due to extrapolation; i.e., the p-air effect is larger.  If one decreases the energy carried by the leading particles (i.e., changes the elasticity) the energy that goes into the shower particles increases.  One would expect the multiplicity to increase also, and this is what happens in the Sibyll model, so this case reduces to the previous one.  The relationship in the Glauber model between the p-p and p-air total cross sections is that a multiplication also occurs here when considering nuclear effects.  In what follows below we use the Glauber model correction for total cross sections.

In this paper we will estimate the uncertainty in extrapolation of total cross section, multiplicity, and elasticity from accelerator energies to 250 TeV in the p-p center of mass, which is about $10^{19.5}$ eV in the lab frame.  We will also make estimates at $10^{17}$ eV, the cosmic ray energy equivalent to 14 TeV in the p-p center of mass,  same as the LHC design energy.  We will then relate these uncertainties to the differences in the models' predictions of $<X_{max}>$.

% %%%%%%%%%%%%%%%%%%%%%%%%%%%%%%%%%%%%%%%%%%%%%%%%%%%%%
\section{Average charged multiplicity:  $<Nch>$}
\label{sec:data}
\begin{figure}[!h]
 \centering
 \includegraphics[width=0.45\textwidth]{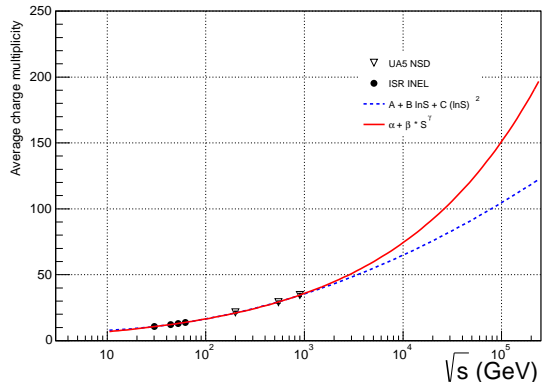}
   \caption{Mean charged multiplicity data for p-p interactions.  Fits to the data are shown to two functions (see text), and the extrapolation to 250 TeV is shown at the end of the fit lines.}
   \label{fig:mult}
\end{figure}

The average charged multiplicity, $<Nch>$, in p$^+$p$^+$ and p$^+$p$^-$ interactions is the same for energies above $\sqrt{s} \sim$~20 GeV, and the highest energy direct measurements were made at the Intersecting Storage Rings~\cite{Breakstone:1983ns} and the CERN Antiproton-Proton collider.  In G.J. Alner  {\it et al.} ~\cite{Alner:1985wj} fits to the charged multiplicity as a function of energy are performed to two functions $<Nch> = A + B~\ln S + C~(\ln S)^2$, and $<Nch> = \alpha + \beta S^{\gamma}$.  Figure~\ref{fig:mult} shows a fit to the data using these two functions. At $\sqrt{s}$ = 14 TeV, the fractional difference is 15$\%$, which results in a 6 gm/cm$^{2}$ uncertainty in $<X_{max}>$.
For the extrapolation to  $\sqrt{s}$ = 250 TeV the fractional difference is 65$\%$, which results in a 25 g/cm$^{2}$ uncertainty in $<X_{max}>$ due to extrapolation.  Of course we do not know the correct function to use in the extrapolation, so these are a lower limits to the uncertainty in  $<X_{max}>$.

%%%%%%%%%%%%%%%%%%%%%%%%%%%%%%%%%%%%%%%%%%%%%%%%%%%%%
%\input{analysis}
\section{Elasticity: }

\begin{figure}[!h]
  \centering
  \subfloat{\includegraphics[width=0.45\textwidth]{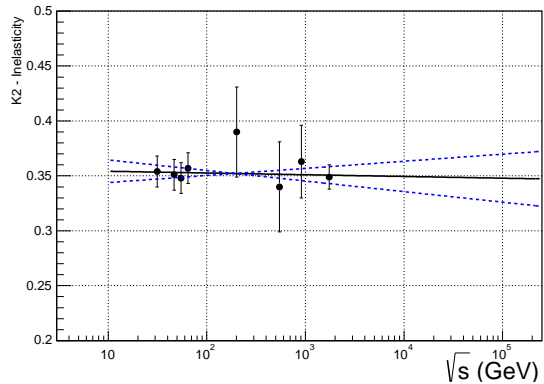}}
  \hfill
  \subfloat{\includegraphics[width=0.45\textwidth]{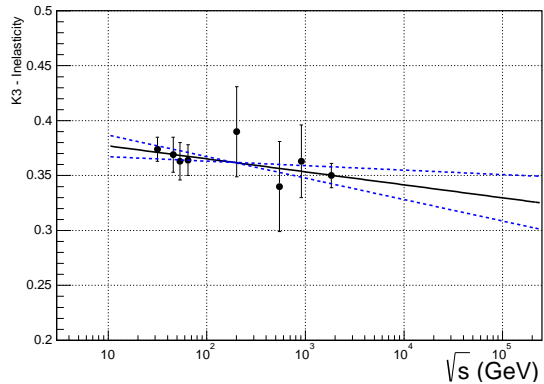}}
  \caption{ Inelasticity of p-p interactions shown as a function of CM energy (GeV).  Two determinations of the inelasticity, called K2 and K3 are shown (see text).  The three lines on each plot represent the central fit (in solid line) and the $\pm$ 1 $\sigma$  (in dotted lines).}
  \label{fig:elas}
\end{figure}

Measurements of inelasticity in pp interactions are somewhat indirect.  They arise from the observation that, $<Nch>$ as a function of $\sqrt{s}$ in e$^+$e$^-$ and pp collisions are very similar, if one lowers the pp $\sqrt{s}$ value; i.e.$<Nch(pp)>(K\sqrt{s}) = <Nch(ee)>(\sqrt{s}) +2$. Here $K=E_{eff}/\sqrt{s}$ where $E_{eff}$ is the effective energy that goes into particle production. The added value of 2 represents the two leading particles in the forward and backward hemispheres, which do not occur in e$^+$e$^-$ jets.  In the review article by J. Grosse-Oetringhaus and K. Reygers~\cite{GrosseOetringhaus:2009kz}, the authors comment that the theoretical basis for this relation is weak.  These authors fit the data as a function of energy to two functions, called K2 and K3, which both have values about 0.35.  There is a slight downward slope as a function of energy, and Figure~\ref{fig:elas} shows fits to the K2 and K3 result as a function of  $\sqrt{s}$ with extrapolation to 250 TeV.  For showers with energy equivalent to $10^{19.5}$ eV (250 TeV),  the fitting function is  the difference between the two inelasticity extrapolations is 20$\%$, yielding a change in elasticity (1-K) of about 10$\%$, so the uncertainty in $<X_{max}>$ is 4 g/cm$^{2}$.  Again the actual function to extrapolate is unknown, so we should interpret this uncertainty as a lower limit.  Figure~\ref{fig:elas} shows the fits to the K2 and K3 functions.

A second review article on this topic by Chliapnikov and Uvarov~\cite{Chliapnikov:1990bc} estimated the value of K to be 0.30. The difference from the fit at low energies by Grosse-Oetringhaus (K $\sim$ 0.35) . This indicates that uncertainties in K are about 16$\%$,  yielding a change in elasticity (1-K) of about 7$\%$,  which corresponds to an uncertainty in $<X_{max}>$ of about 3 g/cm$^{2}$, without extrapolation.  For simplicity we interpret this as the value as the uncertainty in inelasticity at $10^{17}$ eV.

% %%%%%%%%%%%%%%%%%%%%%%%%%%%%%%%%%%%%%%%%%%%%%%%%%%%%%
%\input{sys}

\section{Proton - Air Total Cross Section:}

\label{sec:sys}
\begin{figure}
 \centering
 \includegraphics[width=0.45\textwidth]{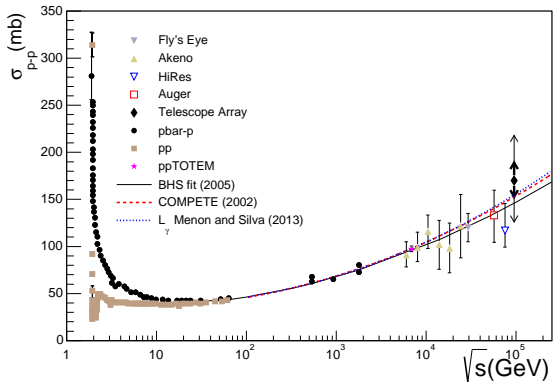}
   \caption{The proton-proton cross section vs. the 
      center of mass energy.  The $\bar{\rm p}$p and the pp data are shown in 
      smaller darker circles and square symbols 
      consecutively~\citep{pp1999}. The recent result from
     LHC is also shown by the star marker~\citep{Antchev:2011vs}. The results by cosmic ray detectors are also included(~\citep{FE1987},~\citep{Honda1992},~\citep{Belov2006},~\citep{Auger2012},~\citep{Abbasi:2015fdr})). The solid line is the QCD inspired fit~\citep{Block:2005pt}, the dashed line is the COMPETE collaboration fit ~\cite{Cudell:2002xe} and the dotted line is the $L_{\gamma}$ fit by Menon and Silva~\cite{Menon:2013vka} .}
   \label{fig:ppmodel}
\end{figure}

Although the uncertainty in $<X_{max}>$ due to the p-air total cross section has been reduced recently, that uncertainty is not zero. The slope of  $\mid \triangle X_{max}$/$\triangle$ f(E) $\mid$ near the cosmic ray energy of $10^{19.5}$ eV  is $\sim$100 g/cm$^{2}$.  Figure~\ref{fig:ppmodel} shows the result of extrapolations to 250 TeV in the center of mass, using three functional fits (the QCD inspired fit by BHS~\cite{Block:2005pt},  the COMPETE collaboration fit~\cite{Cudell:2002xe}, and the $L_{\gamma}$ fit by Menon and Silva ~\cite{Menon:2013vka}).  This figure indicates that the uncertainty in cross section is 8$\%$ at $10^{19.5}$ eV. Using BHS fit~\cite{Block:2005pt} and Glauber model (see Figure 8 in reference~\cite{Abbasi:2015fdr}), in the energy range of interest a change in the p-p total cross section is magnified by a factor of about 1.6 by nuclear effects in the p-air cross section. So a range of uncertainty of 8$\%$ in p-p becomes a range of 13$\%$ in the p-air cross section.  This yields to an uncertainty in $<X_{max}>$ of $\sim$13 g/cm$^{2}$.

for $10^{17}$ eV,  we can make an estimate using recent measurements of the p-p total cross section at LHC, the relation of p-p to p-air cross sections, and the sensitivity in $<X_{max}>$ to the p-air cross section. The cross section measured by the TOTEM~\cite{Antchev:2011vs} experiment at the LHC, at 7 TeV, has an uncertainty of $\pm$ 3$\%$.  So a range of uncertainty of 6$\%$ in p-p becomes a range of 10$\%$ in the p-air cross section.  The slope of  $\mid \triangle X_{max}$/$\triangle$ f(E) $\mid$    near the cosmic ray energy of $10^{17}$ eV is about 44 g/cm$^{2}$, yielding an uncertainty in $<X_{max}>$ of 4 g/cm$^{2}$. 

\section{Uncertainty in $<X_{max}>$ Predictions at $10^{19.5}$ eV:}
Combining the uncertainties in $<X_{max}>$ due to multiplicity, elasticity, and total cross section by simple addition yields an estimate of $\sim$35 g/cm$^{2}$, which is a lower limit to the true uncertainty.  This is about the difference among the five hadronic generator programs in common use today (which is 35 g/cm$^{2}$).  The interpretation of this is that the EPOS-LHC prediction of $<X_{max}>$, which is the highest value of $<X_{max}>$ at $10^{19.5}$ eV, is consistent at the 1$\sigma$ level with the prediction of QGSJet01c, which is the lowest value of $<X_{max}>$, and vice versa.

\section{Uncertainty in $<X_{max}>$ Predictions at $10^{17}$ eV:}
Repeating this procedure for the LHC energy of 14 TeV (equivalent to 1x$10^{17}$ eV in the cosmic ray lab frame) yields a difference among the models of 25 g/cm$^{2}$, and a lower limit to the uncertainty in extrapolation of multiplicity, elasticity, and total cross section of $\sim$ 6 g/cm$^{2}$. The results for four models, and two energies, are tabulated in Table~\ref{tab1}.
%Figure~\ref{fig:ppmodel}  shows the predictions of Corsika and the five hadronic models described above for $<X_{max}>$, and shaded areas representing the extrapolation uncertainties at $10^{17}$ and $10^{19.5}$  eV.

%update numbers
\begin{center}
 \begin{table}   [!hp]
   \begin{tabular}{| p{2cm} | p{2cm} | p{2cm} |}
     \hline
     Model & $<X_{max}>$ uncertainty  $10^{17}$ eV & $<X_{max}>$ uncertainty at $10^{19.5}$ eV \\ \hline
     SIBYLL & 7 gm/cm$^{2}$ & 36 gm/cm$^{2}$ \\ \hline
     QGSJET01 &   6 gm/cm$^{2}$ & 32 gm/cm$^{2}$ \\ \hline
     QGSJETII4 &   6 gm/cm$^{2}$ & 36 gm/cm$^{2}$   \\ \hline
     EPOS-LHC &   6 gm/cm$^{2}$ & 36 gm/cm$^{2}$   \\\hline
 \end{tabular}
    \caption{Results of extrapolations of accelerator measurements. }
   \label{tab1}%
\end{table}
\end{center}

% %%%%%%%%%%%%%%%%%%%%%%%%%%%%%%%%%%%%%%%%%%%%%%%%%%%%%
%\input{pp}
\section{Conclusions}

In recent years there has been considerable improvement in hadronic generator models used in the simulation of cosmic ray showers.  Tuning to the basic cross sections has been improved, and accelerator results at higher energies have been included.  However, knowledge of two of the basic input quantities to hadronic generators, multiplicity and elasticity, have not been improved, and some uncertainty in total cross section remains.  One of the most difficult things to predict is the depth of shower maximum for proton and nuclear primaries.  This is because of the long extrapolation needed to go from energies below $\sim$ one TeV in the center of mass, where accelerator measurements are made, to ~250 TeV where predictions are needed.  By fitting accelerator data on the average number of charged particles produced in proton-proton, proton-antiproton, and electron-positron collisions, fitting to various functions, and extrapolating to higher energies, we have estimated the lower limit on the uncertainty in $<X_{max}>$ from multiplicity, elasticity, and the total cross section.  At the highest energies relevant to cosmic ray studies, the extrapolation on the lower limit of the  uncertainty is approximately equivalent to the difference among the five models considered; i.e., the oldest model, QGSJet01c, is less than 1$\sigma$ different from the newest model, EPOS-LHC.  At the LHC energy itself the lower limit of the extrapolation uncertainty is bit less then the difference among the five models.

% %%%%%%%%%%%%%%%%%%%%%%%%%%%%%%%%%%%%%%%%%%%%%%%%%%%%%
% %\input{acknowledge}
\section{Acknowledgements}
 The authors want to thank  T. Gaisser, P. Sokolsky, J. Belz, and D. Bergman for useful conversations on this topic. We would also like to thank P. Tanguy and  R. Ulrich for providing us with the revision of conex used in this work.

% %%%%%%%%%%%%%%%%%%%%%%%%%%%%%%%%%%%%%%%%%%%%%%%%%%%%%
%\bibliography{paper}{}

\begin{thebibliography}{30}%
\makeatletter
\providecommand \@ifxundefined [1]{%
 \@ifx{#1\undefined}
}%
\providecommand \@ifnum [1]{%
 \ifnum #1\expandafter \@firstoftwo
 \else \expandafter \@secondoftwo
 \fi
}%
\providecommand \@ifx [1]{%
 \ifx #1\expandafter \@firstoftwo
 \else \expandafter \@secondoftwo
 \fi
}%
\providecommand \natexlab [1]{#1}%
\providecommand \enquote  [1]{``#1''}%
\providecommand \bibnamefont  [1]{#1}%
\providecommand \bibfnamefont [1]{#1}%
\providecommand \citenamefont [1]{#1}%
\providecommand \href@noop [0]{\@secondoftwo}%
\providecommand \href [0]{\begingroup \@sanitize@url \@href}%
\providecommand \@href[1]{\@@startlink{#1}\@@href}%
\providecommand \@@href[1]{\endgroup#1\@@endlink}%
\providecommand \@sanitize@url [0]{\catcode `\\12\catcode `\$12\catcode
  `\&12\catcode `\#12\catcode `\^12\catcode `\_12\catcode `\%12\relax}%
\providecommand \@@startlink[1]{}%
\providecommand \@@endlink[0]{}%
\providecommand \url  [0]{\begingroup\@sanitize@url \@url }%
\providecommand \@url [1]{\endgroup\@href {#1}{\urlprefix }}%
\providecommand \urlprefix  [0]{URL }%
\providecommand \Eprint [0]{\href }%
\providecommand \doibase [0]{http://dx.doi.org/}%
\providecommand \selectlanguage [0]{\@gobble}%
\providecommand \bibinfo  [0]{\@secondoftwo}%
\providecommand \bibfield  [0]{\@secondoftwo}%
\providecommand \translation [1]{[#1]}%
\providecommand \BibitemOpen [0]{}%
\providecommand \bibitemStop [0]{}%
\providecommand \bibitemNoStop [0]{.\EOS\space}%
\providecommand \EOS [0]{\spacefactor3000\relax}%
\providecommand \BibitemShut  [1]{\csname bibitem#1\endcsname}%
\let\auto@bib@innerbib\@empty
%</preamble>
\bibitem [{\citenamefont {Abbasi}\ \emph {et~al.}(2010)\citenamefont {Abbasi}
  \emph {et~al.}}]{Abbasi:2009nf}%
  \BibitemOpen
  \bibfield  {author} {\bibinfo {author} {\bibfnamefont {R.~U.}\ \bibnamefont
  {Abbasi}} \emph {et~al.} (\bibinfo {collaboration} {HiRes}),\ }\href
  {\doibase 10.1103/PhysRevLett.104.161101} {\bibfield  {journal} {\bibinfo
  {journal} {Phys. Rev. Lett.}\ }\textbf {\bibinfo {volume} {104}},\ \bibinfo
  {pages} {161101} (\bibinfo {year} {2010})},\ \Eprint
  {http://arxiv.org/abs/0910.4184} {arXiv:0910.4184 [astro-ph.HE]} \BibitemShut
  {NoStop}%
%%CITATION = ARXIV:0910.4184;%%
\bibitem [{\citenamefont {Abbasi}\ \emph
  {et~al.}(2015{\natexlab{a}})\citenamefont {Abbasi} \emph
  {et~al.}}]{Abbasi:2015bha}%
  \BibitemOpen
  \bibfield  {author} {\bibinfo {author} {\bibfnamefont {R.~U.}\ \bibnamefont
  {Abbasi}} \emph {et~al.},\ }\href {\doibase
  10.1016/j.astropartphys.2015.02.008} {\bibfield  {journal} {\bibinfo
  {journal} {Astropart. Phys.}\ }\textbf {\bibinfo {volume} {68}},\ \bibinfo
  {pages} {27} (\bibinfo {year} {2015}{\natexlab{a}})}\BibitemShut {NoStop}%
%%CITATION = APHYE,68,27;%%
\bibitem [{\citenamefont {Abraham}\ \emph {et~al.}(2010)\citenamefont {Abraham}
  \emph {et~al.}}]{Abraham:2010yv}%
  \BibitemOpen
  \bibfield  {author} {\bibinfo {author} {\bibfnamefont {J.}~\bibnamefont
  {Abraham}} \emph {et~al.} (\bibinfo {collaboration} {Pierre Auger}),\ }\href
  {\doibase 10.1103/PhysRevLett.104.091101} {\bibfield  {journal} {\bibinfo
  {journal} {Phys. Rev. Lett.}\ }\textbf {\bibinfo {volume} {104}},\ \bibinfo
  {pages} {091101} (\bibinfo {year} {2010})},\ \Eprint
  {http://arxiv.org/abs/1002.0699} {arXiv:1002.0699 [astro-ph.HE]} \BibitemShut
  {NoStop}%
%%CITATION = ARXIV:1002.0699;%%
\bibitem [{\citenamefont {Aab}\ \emph {et~al.}(2014)\citenamefont {Aab} \emph
  {et~al.}}]{Aab:2014aea}%
  \BibitemOpen
  \bibfield  {author} {\bibinfo {author} {\bibfnamefont {A.}~\bibnamefont
  {Aab}} \emph {et~al.} (\bibinfo {collaboration} {Pierre Auger}),\ }\href
  {\doibase 10.1103/PhysRevD.90.122006} {\bibfield  {journal} {\bibinfo
  {journal} {Phys. Rev.}\ }\textbf {\bibinfo {volume} {D90}},\ \bibinfo {pages}
  {122006} (\bibinfo {year} {2014})},\ \Eprint {http://arxiv.org/abs/1409.5083}
  {arXiv:1409.5083 [astro-ph.HE]} \BibitemShut {NoStop}%
%%CITATION = ARXIV:1409.5083;%%
\bibitem [{\citenamefont {Abbasi}\ \emph
  {et~al.}(2015{\natexlab{b}})\citenamefont {Abbasi} \emph
  {et~al.}}]{Abbasi:2015xga}%
  \BibitemOpen
  \bibfield  {author} {\bibinfo {author} {\bibfnamefont {R.}~\bibnamefont
  {Abbasi}} \emph {et~al.} (\bibinfo {collaboration} {Pierre Auger, Telescope
  Array}),\ }in\ \href
  {http://inspirehep.net/record/1356235/files/arXiv:1503.07540.pdf} {\emph
  {\bibinfo {booktitle} {{2014 Conference on Ultrahigh Energy Cosmic Rays
  (UHECR2014) Springdale, USA, October 12-15, 2014}}}}\ (\bibinfo {year}
  {2015})\ \Eprint {http://arxiv.org/abs/1503.07540} {arXiv:1503.07540
  [astro-ph.HE]} \BibitemShut {NoStop}%
%%CITATION = ARXIV:1503.07540;%%
\bibitem [{\citenamefont {Heck}\ \emph {et~al.}(1998)\citenamefont {Heck},
  \citenamefont {Schatz}, \citenamefont {Thouw}, \citenamefont {Knapp},\ and\
  \citenamefont {Capdevielle}}]{Heck:1998vt}%
  \BibitemOpen
  \bibfield  {author} {\bibinfo {author} {\bibfnamefont {D.}~\bibnamefont
  {Heck}}, \bibinfo {author} {\bibfnamefont {G.}~\bibnamefont {Schatz}},
  \bibinfo {author} {\bibfnamefont {T.}~\bibnamefont {Thouw}}, \bibinfo
  {author} {\bibfnamefont {J.}~\bibnamefont {Knapp}}, \ and\ \bibinfo {author}
  {\bibfnamefont {J.~N.}\ \bibnamefont {Capdevielle}},\ }\href@noop {} {\
  (\bibinfo {year} {1998})}\BibitemShut {NoStop}%
%%CITATION = FZKA-6019;%%
\bibitem [{\citenamefont {Bergmann}\ \emph {et~al.}(2007)\citenamefont
  {Bergmann} \emph {et~al.}}]{Bergmann:2006yz}%
  \BibitemOpen
  \bibfield  {author} {\bibinfo {author} {\bibfnamefont {T.}~\bibnamefont
  {Bergmann}} \emph {et~al.},\ }\href@noop {} {\bibfield  {journal} {\bibinfo
  {journal} {Astropart. Phys.}\ }\textbf {\bibinfo {volume} {26}},\ \bibinfo
  {pages} {420} (\bibinfo {year} {2007})},\ \Eprint
  {http://arxiv.org/abs/astro-ph/0606564} {astro-ph/0606564} \BibitemShut
  {NoStop}%
%%CITATION = ASTRO-PH/0606564;%%
\bibitem [{\citenamefont {Pierog}\ \emph {et~al.}(2006)\citenamefont {Pierog}
  \emph {et~al.}}]{Pierog:2004re}%
  \BibitemOpen
  \bibfield  {author} {\bibinfo {author} {\bibfnamefont {T.}~\bibnamefont
  {Pierog}} \emph {et~al.},\ }\href@noop {} {\bibfield  {journal} {\bibinfo
  {journal} {Nucl. Phys. Proc. Suppl.}\ }\textbf {\bibinfo {volume} {151}},\
  \bibinfo {pages} {159} (\bibinfo {year} {2006})},\ \Eprint
  {http://arxiv.org/abs/astro-ph/0411260} {astro-ph/0411260} \BibitemShut
  {NoStop}%
%%CITATION = ASTRO-PH/0411260;%%
\bibitem [{\citenamefont {Bossard}\ \emph {et~al.}(2001)\citenamefont {Bossard}
  \emph {et~al.}}]{Bossard:2000jh}%
  \BibitemOpen
  \bibfield  {author} {\bibinfo {author} {\bibfnamefont {G.}~\bibnamefont
  {Bossard}} \emph {et~al.},\ }\href@noop {} {\bibfield  {journal} {\bibinfo
  {journal} {Phys. Rev.}\ }\textbf {\bibinfo {volume} {D63}},\ \bibinfo {pages}
  {054030} (\bibinfo {year} {2001})},\ \Eprint
  {http://arxiv.org/abs/hep-ph/0009119} {hep-ph/0009119} \BibitemShut {NoStop}%
%%CITATION = HEP-PH/0009119;%%
\bibitem [{\citenamefont {Kalmykov}\ \emph {et~al.}(1997)\citenamefont
  {Kalmykov}, \citenamefont {Ostapchenko},\ and\ \citenamefont
  {Pavlov}}]{Kalmykov:1997te}%
  \BibitemOpen
  \bibfield  {author} {\bibinfo {author} {\bibfnamefont {N.~N.}\ \bibnamefont
  {Kalmykov}}, \bibinfo {author} {\bibfnamefont {S.~S.}\ \bibnamefont
  {Ostapchenko}}, \ and\ \bibinfo {author} {\bibfnamefont {A.~I.}\ \bibnamefont
  {Pavlov}},\ }\bibfield  {booktitle} {\emph {\bibinfo {booktitle}
  {{Proceedings, 9th International Symposium on Very High Energy Cosmic Ray
  Interactions (ISVHECRI 1996)}}},\ }\href {\doibase
  10.1016/S0920-5632(96)00846-8} {\bibfield  {journal} {\bibinfo  {journal}
  {Nucl. Phys. Proc. Suppl.}\ }\textbf {\bibinfo {volume} {52}},\ \bibinfo
  {pages} {17} (\bibinfo {year} {1997})}\BibitemShut {NoStop}%
%%CITATION = NUPHZ,52,17;%%
\bibitem [{\citenamefont {Ostapchenko}(2006)}]{Ostapchenko:2005nj}%
  \BibitemOpen
  \bibfield  {author} {\bibinfo {author} {\bibfnamefont {S.}~\bibnamefont
  {Ostapchenko}},\ }\href {\doibase 10.1103/PhysRevD.74.014026} {\bibfield
  {journal} {\bibinfo  {journal} {Phys. Rev.}\ }\textbf {\bibinfo {volume}
  {D74}},\ \bibinfo {pages} {014026} (\bibinfo {year} {2006})},\ \Eprint
  {http://arxiv.org/abs/hep-ph/0505259} {arXiv:hep-ph/0505259 [hep-ph]}
  \BibitemShut {NoStop}%
%%CITATION = HEP-PH/0505259;%%
\bibitem [{\citenamefont {Ostapchenko}(2011)}]{Ostapchenko:2010vb}%
  \BibitemOpen
  \bibfield  {author} {\bibinfo {author} {\bibfnamefont {S.}~\bibnamefont
  {Ostapchenko}},\ }\href {\doibase 10.1103/PhysRevD.83.014018} {\bibfield
  {journal} {\bibinfo  {journal} {Phys. Rev.}\ }\textbf {\bibinfo {volume}
  {D83}},\ \bibinfo {pages} {014018} (\bibinfo {year} {2011})},\ \Eprint
  {http://arxiv.org/abs/1010.1869} {arXiv:1010.1869 [hep-ph]} \BibitemShut
  {NoStop}%
%%CITATION = ARXIV:1010.1869;%%
\bibitem [{\citenamefont {Fletcher}\ \emph {et~al.}(1994)\citenamefont
  {Fletcher}, \citenamefont {Gaisser}, \citenamefont {Lipari},\ and\
  \citenamefont {Stanev}}]{Fletcher:1994bd}%
  \BibitemOpen
  \bibfield  {author} {\bibinfo {author} {\bibfnamefont {R.~S.}\ \bibnamefont
  {Fletcher}}, \bibinfo {author} {\bibfnamefont {T.~K.}\ \bibnamefont
  {Gaisser}}, \bibinfo {author} {\bibfnamefont {P.}~\bibnamefont {Lipari}}, \
  and\ \bibinfo {author} {\bibfnamefont {T.}~\bibnamefont {Stanev}},\ }\href
  {\doibase 10.1103/PhysRevD.50.5710} {\bibfield  {journal} {\bibinfo
  {journal} {Phys. Rev.}\ }\textbf {\bibinfo {volume} {D50}},\ \bibinfo {pages}
  {5710} (\bibinfo {year} {1994})}\BibitemShut {NoStop}%
%%CITATION = PHRVA,D50,5710;%%
\bibitem [{\citenamefont {Engel}\ \emph {et~al.}(1999)\citenamefont {Engel},
  \citenamefont {Gaisser}, \citenamefont {Stanev},\ and\ \citenamefont
  {Lipari}}]{Engel:1999db}%
  \BibitemOpen
  \bibfield  {author} {\bibinfo {author} {\bibfnamefont {R.}~\bibnamefont
  {Engel}}, \bibinfo {author} {\bibfnamefont {T.~K.}\ \bibnamefont {Gaisser}},
  \bibinfo {author} {\bibfnamefont {T.}~\bibnamefont {Stanev}}, \ and\ \bibinfo
  {author} {\bibfnamefont {P.}~\bibnamefont {Lipari}},\ }in\ \href
  {http://krusty.physics.utah.edu/~icrc1999/root/vol1/h2_5_03.pdf} {\emph
  {\bibinfo {booktitle} {{Proceedings, 26th International Cosmic Ray
  Conference, August 17-25, 1999, Salt Lake City}}}},\ Vol.~\bibinfo {volume}
  {1}\ (\bibinfo {year} {1999})\ pp.\ \bibinfo {pages} {415--418}\BibitemShut
  {NoStop}%
%%CITATION = INSPIRE-512007;%%
\bibitem [{\citenamefont {Pierog}\ \emph {et~al.}(2015)\citenamefont {Pierog},
  \citenamefont {Karpenko}, \citenamefont {Katzy}, \citenamefont {Yatsenko},\
  and\ \citenamefont {Werner}}]{Pierog:2013ria}%
  \BibitemOpen
  \bibfield  {author} {\bibinfo {author} {\bibfnamefont {T.}~\bibnamefont
  {Pierog}}, \bibinfo {author} {\bibfnamefont {I.}~\bibnamefont {Karpenko}},
  \bibinfo {author} {\bibfnamefont {J.~M.}\ \bibnamefont {Katzy}}, \bibinfo
  {author} {\bibfnamefont {E.}~\bibnamefont {Yatsenko}}, \ and\ \bibinfo
  {author} {\bibfnamefont {K.}~\bibnamefont {Werner}},\ }\href {\doibase
  10.1103/PhysRevC.92.034906} {\bibfield  {journal} {\bibinfo  {journal} {Phys.
  Rev.}\ }\textbf {\bibinfo {volume} {C92}},\ \bibinfo {pages} {034906}
  (\bibinfo {year} {2015})},\ \Eprint {http://arxiv.org/abs/1306.0121}
  {arXiv:1306.0121 [hep-ph]} \BibitemShut {NoStop}%
%%CITATION = ARXIV:1306.0121;%%
\bibitem [{\citenamefont {Ulrich}\ \emph {et~al.}(2011)\citenamefont {Ulrich},
  \citenamefont {Engel},\ and\ \citenamefont {Unger}}]{Ulrich:2010rg}%
  \BibitemOpen
  \bibfield  {author} {\bibinfo {author} {\bibfnamefont {R.}~\bibnamefont
  {Ulrich}}, \bibinfo {author} {\bibfnamefont {R.}~\bibnamefont {Engel}}, \
  and\ \bibinfo {author} {\bibfnamefont {M.}~\bibnamefont {Unger}},\ }\href
  {\doibase 10.1103/PhysRevD.83.054026} {\bibfield  {journal} {\bibinfo
  {journal} {Phys. Rev.}\ }\textbf {\bibinfo {volume} {D83}},\ \bibinfo {pages}
  {054026} (\bibinfo {year} {2011})},\ \Eprint {http://arxiv.org/abs/1010.4310}
  {arXiv:1010.4310 [hep-ph]} \BibitemShut {NoStop}%
%%CITATION = ARXIV:1010.4310;%%
\bibitem [{\citenamefont {Breakstone}\ \emph {et~al.}(1984)\citenamefont
  {Breakstone} \emph {et~al.}}]{Breakstone:1983ns}%
  \BibitemOpen
  \bibfield  {author} {\bibinfo {author} {\bibfnamefont {A.}~\bibnamefont
  {Breakstone}} \emph {et~al.} (\bibinfo {collaboration}
  {Ames-Bologna-CERN-Dortmund-Heidelberg-Warsaw}),\ }\href {\doibase
  10.1103/PhysRevD.30.528} {\bibfield  {journal} {\bibinfo  {journal} {Phys.
  Rev.}\ }\textbf {\bibinfo {volume} {D30}},\ \bibinfo {pages} {528} (\bibinfo
  {year} {1984})}\BibitemShut {NoStop}%
%%CITATION = PHRVA,D30,528;%%
\bibitem [{\citenamefont {Alner}\ \emph {et~al.}(1986)\citenamefont {Alner}
  \emph {et~al.}}]{Alner:1985wj}%
  \BibitemOpen
  \bibfield  {author} {\bibinfo {author} {\bibfnamefont {G.~J.}\ \bibnamefont
  {Alner}} \emph {et~al.} (\bibinfo {collaboration} {UA5}),\ }\href {\doibase
  10.1016/0370-2693(86)91304-3} {\bibfield  {journal} {\bibinfo  {journal}
  {Phys. Lett.}\ }\textbf {\bibinfo {volume} {B167}},\ \bibinfo {pages} {476}
  (\bibinfo {year} {1986})}\BibitemShut {NoStop}%
%%CITATION = PHLTA,B167,476;%%
\bibitem [{\citenamefont {Grosse-Oetringhaus}\ and\ \citenamefont
  {Reygers}(2010)}]{GrosseOetringhaus:2009kz}%
  \BibitemOpen
  \bibfield  {author} {\bibinfo {author} {\bibfnamefont {J.~F.}\ \bibnamefont
  {Grosse-Oetringhaus}}\ and\ \bibinfo {author} {\bibfnamefont
  {K.}~\bibnamefont {Reygers}},\ }\href {\doibase
  10.1088/0954-3899/37/8/083001} {\bibfield  {journal} {\bibinfo  {journal} {J.
  Phys.}\ }\textbf {\bibinfo {volume} {G37}},\ \bibinfo {pages} {083001}
  (\bibinfo {year} {2010})},\ \Eprint {http://arxiv.org/abs/0912.0023}
  {arXiv:0912.0023 [hep-ex]} \BibitemShut {NoStop}%
%%CITATION = ARXIV:0912.0023;%%
\bibitem [{\citenamefont {Chliapnikov}\ and\ \citenamefont
  {Uvarov}(1990)}]{Chliapnikov:1990bc}%
  \BibitemOpen
  \bibfield  {author} {\bibinfo {author} {\bibfnamefont {P.~V.}\ \bibnamefont
  {Chliapnikov}}\ and\ \bibinfo {author} {\bibfnamefont {V.~A.}\ \bibnamefont
  {Uvarov}},\ }\href {\doibase 10.1016/0370-2693(90)90252-2} {\bibfield
  {journal} {\bibinfo  {journal} {Phys. Lett.}\ }\textbf {\bibinfo {volume}
  {B251}},\ \bibinfo {pages} {192} (\bibinfo {year} {1990})}\BibitemShut
  {NoStop}%
%%CITATION = PHLTA,B251,192;%%
\bibitem [{\citenamefont {Block}\ and\ \citenamefont
  {Halzen}(2005)}]{Block:2005pt}%
  \BibitemOpen
  \bibfield  {author} {\bibinfo {author} {\bibfnamefont {M.~M.}\ \bibnamefont
  {Block}}\ and\ \bibinfo {author} {\bibfnamefont {F.}~\bibnamefont {Halzen}},\
  }\href {\doibase 10.1103/PhysRevD.72.036006, 10.1103/PhysRevD.72.039902}
  {\bibfield  {journal} {\bibinfo  {journal} {Phys. Rev.}\ }\textbf {\bibinfo
  {volume} {D72}},\ \bibinfo {pages} {036006} (\bibinfo {year} {2005})},\
  \bibinfo {note} {[Erratum: Phys. Rev.D72,039902(2005)]},\ \Eprint
  {http://arxiv.org/abs/hep-ph/0506031} {arXiv:hep-ph/0506031 [hep-ph]}
  \BibitemShut {NoStop}%
%%CITATION = HEP-PH/0506031;%%
\bibitem [{\citenamefont {Cudell}\ \emph {et~al.}(2002)\citenamefont {Cudell},
  \citenamefont {Ezhela}, \citenamefont {Gauron}, \citenamefont {Kang},
  \citenamefont {Kuyanov}, \citenamefont {Lugovsky}, \citenamefont {Martynov},
  \citenamefont {Nicolescu}, \citenamefont {Razuvaev},\ and\ \citenamefont
  {Tkachenko}}]{Cudell:2002xe}%
  \BibitemOpen
  \bibfield  {author} {\bibinfo {author} {\bibfnamefont {J.~R.}\ \bibnamefont
  {Cudell}}, \bibinfo {author} {\bibfnamefont {V.~V.}\ \bibnamefont {Ezhela}},
  \bibinfo {author} {\bibfnamefont {P.}~\bibnamefont {Gauron}}, \bibinfo
  {author} {\bibfnamefont {K.}~\bibnamefont {Kang}}, \bibinfo {author}
  {\bibfnamefont {{\relax Yu}.~V.}\ \bibnamefont {Kuyanov}}, \bibinfo {author}
  {\bibfnamefont {S.~B.}\ \bibnamefont {Lugovsky}}, \bibinfo {author}
  {\bibfnamefont {E.}~\bibnamefont {Martynov}}, \bibinfo {author}
  {\bibfnamefont {B.}~\bibnamefont {Nicolescu}}, \bibinfo {author}
  {\bibfnamefont {E.~A.}\ \bibnamefont {Razuvaev}}, \ and\ \bibinfo {author}
  {\bibfnamefont {N.~P.}\ \bibnamefont {Tkachenko}} (\bibinfo {collaboration}
  {COMPETE}),\ }\href {\doibase 10.1103/PhysRevLett.89.201801} {\bibfield
  {journal} {\bibinfo  {journal} {Phys. Rev. Lett.}\ }\textbf {\bibinfo
  {volume} {89}},\ \bibinfo {pages} {201801} (\bibinfo {year} {2002})},\
  \Eprint {http://arxiv.org/abs/hep-ph/0206172} {arXiv:hep-ph/0206172 [hep-ph]}
  \BibitemShut {NoStop}%
%%CITATION = HEP-PH/0206172;%%
\bibitem [{\citenamefont {Menon}\ and\ \citenamefont
  {Silva}(2013)}]{Menon:2013vka}%
  \BibitemOpen
  \bibfield  {author} {\bibinfo {author} {\bibfnamefont {M.~J.}\ \bibnamefont
  {Menon}}\ and\ \bibinfo {author} {\bibfnamefont {P.~V. R.~G.}\ \bibnamefont
  {Silva}},\ }\href {\doibase 10.1088/0954-3899/41/1/019501,
  10.1088/0954-3899/40/12/125001} {\bibfield  {journal} {\bibinfo  {journal}
  {J. Phys.}\ }\textbf {\bibinfo {volume} {G40}},\ \bibinfo {pages} {125001}
  (\bibinfo {year} {2013})},\ \bibinfo {note} {[Erratum: J.
  Phys.G41,019501(2014)]},\ \Eprint {http://arxiv.org/abs/1305.2947}
  {arXiv:1305.2947 [hep-ph]} \BibitemShut {NoStop}%
%%CITATION = ARXIV:1305.2947;%%
\bibitem [{\citenamefont {Avila}\ \emph {et~al.}(1999)\citenamefont {Avila}
  \emph {et~al.}}]{pp1999}%
  \BibitemOpen
  \bibfield  {author} {\bibinfo {author} {\bibfnamefont {C.}~\bibnamefont
  {Avila}} \emph {et~al.},\ }\href {\doibase 10.1016/S0370-2693(98)01421-X}
  {\bibfield  {journal} {\bibinfo  {journal} {Phys.Lett. B}\ }\textbf {\bibinfo
  {volume} {445}},\ \bibinfo {pages} {419–422} (\bibinfo {year}
  {1999})}\BibitemShut {NoStop}%
\bibitem [{\citenamefont {Antchev}\ \emph {et~al.}(2011)\citenamefont {Antchev}
  \emph {et~al.}}]{Antchev:2011vs}%
  \BibitemOpen
  \bibfield  {author} {\bibinfo {author} {\bibfnamefont {G.}~\bibnamefont
  {Antchev}} \emph {et~al.},\ }\href {\doibase 10.1209/0295-5075/96/21002}
  {\bibfield  {journal} {\bibinfo  {journal} {Europhys. Lett.}\ }\textbf
  {\bibinfo {volume} {96}},\ \bibinfo {pages} {21002} (\bibinfo {year}
  {2011})},\ \Eprint {http://arxiv.org/abs/1110.1395} {arXiv:1110.1395
  [hep-ex]} \BibitemShut {NoStop}%
%%CITATION = ARXIV:1110.1395;%%
\bibitem [{\citenamefont {Baltrusaitis}\ \emph {et~al.}(1984)\citenamefont
  {Baltrusaitis}, \citenamefont {Cassiday}, \citenamefont {Elbert},
  \citenamefont {Gerhardy}, \citenamefont {Ko} \emph {et~al.}}]{FE1987}%
  \BibitemOpen
  \bibfield  {author} {\bibinfo {author} {\bibfnamefont {R.}~\bibnamefont
  {Baltrusaitis}}, \bibinfo {author} {\bibfnamefont {G.}~\bibnamefont
  {Cassiday}}, \bibinfo {author} {\bibfnamefont {J.}~\bibnamefont {Elbert}},
  \bibinfo {author} {\bibfnamefont {P.}~\bibnamefont {Gerhardy}}, \bibinfo
  {author} {\bibfnamefont {S.}~\bibnamefont {Ko}},  \emph {et~al.},\ }\href
  {\doibase 10.1103/PhysRevLett.52.1380} {\bibfield  {journal} {\bibinfo
  {journal} {Phys.Rev.Lett.}\ }\textbf {\bibinfo {volume} {52}},\ \bibinfo
  {pages} {1380} (\bibinfo {year} {1984})}\BibitemShut {NoStop}%
%%CITATION = PRLTA,52,1380;%%
\bibitem [{\citenamefont {Honda}\ \emph {et~al.}(1993)\citenamefont {Honda},
  \citenamefont {Nagano}, \citenamefont {Tonwar}, \citenamefont {Kasahara},
  \citenamefont {Hara} \emph {et~al.}}]{Honda1992}%
  \BibitemOpen
  \bibfield  {author} {\bibinfo {author} {\bibfnamefont {M.}~\bibnamefont
  {Honda}}, \bibinfo {author} {\bibfnamefont {M.}~\bibnamefont {Nagano}},
  \bibinfo {author} {\bibfnamefont {S.}~\bibnamefont {Tonwar}}, \bibinfo
  {author} {\bibfnamefont {K.}~\bibnamefont {Kasahara}}, \bibinfo {author}
  {\bibfnamefont {T.}~\bibnamefont {Hara}},  \emph {et~al.},\ }\href {\doibase
  10.1103/PhysRevLett.70.525} {\bibfield  {journal} {\bibinfo  {journal}
  {Phys.Rev.Lett.}\ }\textbf {\bibinfo {volume} {70}},\ \bibinfo {pages} {525}
  (\bibinfo {year} {1993})}\BibitemShut {NoStop}%
%%CITATION = PRLTA,70,525;%%
\bibitem [{\citenamefont {Belov}(2006)}]{Belov2006}%
  \BibitemOpen
  \bibfield  {author} {\bibinfo {author} {\bibfnamefont {K.}~\bibnamefont
  {Belov}} (\bibinfo {collaboration} {HiRes Collaboration}),\ }\href {\doibase
  10.1016/j.nuclphysbps.2005.07.035} {\bibfield  {journal} {\bibinfo  {journal}
  {Nucl.Phys.Proc.Suppl.}\ }\textbf {\bibinfo {volume} {151}},\ \bibinfo
  {pages} {197} (\bibinfo {year} {2006})}\BibitemShut {NoStop}%
%%CITATION = NUPHZ,151,197;%%
\bibitem [{\citenamefont {Abreu}\ \emph {et~al.}(2012)\citenamefont {Abreu}
  \emph {et~al.}}]{Auger2012}%
  \BibitemOpen
  \bibfield  {author} {\bibinfo {author} {\bibfnamefont {P.}~\bibnamefont
  {Abreu}} \emph {et~al.} (\bibinfo {collaboration} {Pierre Auger
  Collaboration}),\ }\href {\doibase 10.1103/PhysRevLett.109.062002} {\bibfield
   {journal} {\bibinfo  {journal} {Phys.Rev.Lett.}\ }\textbf {\bibinfo {volume}
  {109}},\ \bibinfo {pages} {062002} (\bibinfo {year} {2012})},\ \Eprint
  {http://arxiv.org/abs/1208.1520} {arXiv:1208.1520 [hep-ex]} \BibitemShut
  {NoStop}%
%%CITATION = ARXIV:1208.1520;%%
\bibitem [{\citenamefont {Abbasi}\ \emph
  {et~al.}(2015{\natexlab{c}})\citenamefont {Abbasi} \emph
  {et~al.}}]{Abbasi:2015fdr}%
  \BibitemOpen
  \bibfield  {author} {\bibinfo {author} {\bibfnamefont {R.~U.}\ \bibnamefont
  {Abbasi}} \emph {et~al.} (\bibinfo {collaboration} {Telescope Array}),\
  }\href {\doibase 10.1103/PhysRevD.92.032007} {\bibfield  {journal} {\bibinfo
  {journal} {Phys. Rev.}\ }\textbf {\bibinfo {volume} {D92}},\ \bibinfo {pages}
  {032007} (\bibinfo {year} {2015}{\natexlab{c}})},\ \Eprint
  {http://arxiv.org/abs/1505.01860} {arXiv:1505.01860 [astro-ph.HE]}
  \BibitemShut {NoStop}%
%%CITATION = ARXIV:1505.01860;%%
\end{thebibliography}

\end{document}